\documentclass[aps, twocolumn, superscriptaddress, showpacs, longbibliography]{revtex4-1}

\usepackage{amsmath, amsfonts, amssymb, amsthm, dsfont}
\usepackage{hyperref}
\usepackage{physics}
\usepackage{tabularx}
\usepackage{tikz}
\usepackage{usefulnotations}

\begin{document}
	\title{Non-Hermitian Rayleigh-Schr\"{o}dinger-like Perturbation Theory at Exceptional Point}

	\author{Wei-Ming Chen}
	\affiliation{Department of Physics, National Sun Yat-sen University, Kaohsiung 80424, Taiwan}
	\affiliation{Center for Theoretical and Computational Physics, National Sun Yat-sen University, Kaohsiung 80424, Taiwan}
	\author{Chia-Yi Ju}
	\email{chiayiju@mail.nsysu.edu.tw}
	\affiliation{Department of Physics, National Sun Yat-sen University, Kaohsiung 80424, Taiwan}
	\affiliation{Center for Theoretical and Computational Physics, National Sun Yat-sen University, Kaohsiung 80424, Taiwan}
	\affiliation{Physics Division, National Center for Theoretical Sciences, Taipei 106319, Taiwan}

	\begin{abstract}
		We develop a Rayleigh--Schr\"{o}dinger-like perturbation theory for non-Hermitian quantum systems at an exceptional point of order $N$\,. Working in the Jordan basis of the unperturbed Hamiltonian and employing a Puiseux expansion of the perturbed eigenvalues and eigenstates, we derive explicit recursion relations for the expansion coefficients. The corrections to the unperturbed eigenvalue in the Puiseux expansion govern the splitting near the exceptional point; the first two are obtained iteratively in two equivalent forms. One is given in terms of the perturbation Hamiltonian in the Jordan basis, and the other in terms of the generator that drives the eigenvalue evolution with respect to the perturbation. The latter constitutes the exceptional-point counterpart of a geometric perturbation method recently developed for the non-exceptional-point regime. Both representations are verified explicitly for the $N = 2$ and $N = 3$ cases.
	\end{abstract}

	\maketitle
	\newpage

	\section{Introduction}
	\label{sec:intro}

Non-Hermitian quantum mechanics has emerged as a rich and active frontier, revealing physical phenomena that fall outside the conventional Hermitian framework. Over the past decades, these phenomena have been experimentally observed in systems ranging from classical microwave cavities~\cite{Dembowski2001, Dembowski2003, Dembowski2004} and coupled optical microcavities~\cite{Peng2014} to solid-state quantum spin architectures~\cite{Wu2019}. At the heart of these diverse experimental realizations lies the profound theoretical concept of \textit{exceptional points} (EPs)~
\cite{Bender1998,Mostafazadeh2004,Bender2007,Mostafazadeh2010,Brody2013,Wiersig2014,Mostafazadeh2018,Ju2019,Ju2022,Arkhipov2023,Lai2024}, which are parameter singularities where two or more eigenvalues and their corresponding eigenvectors coalesce simultaneously~\cite{Kato1966,Heiss2000,Berry2004,Heiss2004,MiriAlu2019}. At an order-$N$ EP, the Hamiltonian is no longer diagonalizable but instead admits a single Jordan block of size $N$\,, with the sole eigenvector being self-orthogonal under the biorthogonal inner product. Perturbing away from the EP, the resulting non-analytic behavior of the eigenvalues makes a detailed quantitative description particularly relevant.

To this end, perturbation theory has long been an indispensable tool for understanding quantum systems near exactly solvable limits. The Rayleigh--Schr\"{o}dinger (RS) perturbation theory~\cite{Schroedinger1926} provides systematic corrections to eigenvalues and eigenstates when the Hamiltonian is slightly deformed away from a reference point. However, this framework breaks down at an EP: the perturbed eigenvalue is no longer analytic in the perturbation parameter $\epsilon$\,, splitting instead as fractional powers rather than as an integer power series. The proper mathematical setting is the theory of Puiseux series~\cite{Kato1966}, which represents the perturbed eigenvalue as a convergent series in fractional powers of $\epsilon$\,. Under a generic condition on the perturbation, the leading corrections are governed by the perturbation and the generalized eigenvectors of the unperturbed Hamiltonian~\cite{Lidskii1966,VishikLyusternik1960,Moro1997,Seyranian2003}. For a single Jordan block, this genericity condition can be stated explicitly, and the Puiseux coefficients admit explicit recursive formulas to arbitrary order~\cite{Welters2011} (see also~\cite{Mailybaev2006,Znojil2020a}).

		A complementary line of development concerns a geometric formulation of (non-)Hermitian quantum mechanics~\cite{Ju2024,Ju2026,Tu2023}, in which a physical state is parallel-transported in both the time $t$ and a physical parameter $q$. In this picture, the Hamiltonian $H$ is the $t$-evolution generator, while the $q$-evolution is generated by $K = K_1 t + K_0$\,. The time-independent generators $K_1$ and $K_0$ dictate the evolution of the eigenvalues and eigenstates, respectively. Within this geometric framework, a systematic non-Hermitian generalization of RS perturbation theory was recently developed in~\cite{original_paper}, where the perturbed eigenvalue is expressed order by order in terms of the eigenvalue-evolution generator, and the method is shown to reduce to the standard Hermitian RS theory in the appropriate limit. Yet, as explicitly noted in~\cite{original_paper}, this geometric perturbation theory breaks down at an EP: the biorthogonal eigenbasis degenerates, the metric of the Hilbert space becomes singular, and the standard power-series expansion of the eigenvalues fails. Extending the geometric framework to the EP regime has yet to be addressed.

		In this work we fill this gap by developing a systematic Rayleigh--Schr\"{o}dinger-like perturbation theory directly at an EP, exploiting the Jordan basis structure of the Hamiltonian and employing a Puiseux expansion in $\epsilon^{1/N}$\,. We derive explicit recursion relations for the corrections to the unperturbed eigenvalue and eigenstate. Solving them iteratively allows us to obtain the first two corrections, $\p{h}{1/N}$ and $\p{h}{2/N}$, representing the coefficients of $\epsilon^{1/N}$ and $\epsilon^{2/N}$ in the Puiseux expansion. The results are expressed in two complementary representations: in terms of the perturbation matrix $\p{H}{1}$ [Eqs.~\eqref{h1NHp} and \eqref{eq:h2N_final}], and in terms of $\p{K}{0}_1$,  the zeroth-order coefficient of the Laurent expansion of $K_1$ [Eqs.~\eqref{eq:h1N_K1} and \eqref{eq:h2N_K1}]. The latter constitutes the EP analogue of the non-EP result of~\cite{original_paper}: at an order-$N$ EP, the $N$-th power of the leading correction is determined by a single matrix element of $\p{K}{0}_1$, while the sub-leading correction is given by a linear combination of matrix elements of $\p{K}{0}_1$\,. Both representations are verified for the $N = 2$ and $N = 3$ cases. Beyond these specific cases, this perturbative procedure provides a Rayleigh--Schr\"{o}dinger-like way to reproduce the leading correction obtained in the classical Lidskii--Vishik--Lyusternik perturbation theory~\cite{Lidskii1966,VishikLyusternik1960,Moro1997,Seyranian2003} and the recursion of~\cite{Welters2011}.

		The remainder of this paper is organized as follows. In Sec.~\ref{sec:main}, we introduce the physical setting and establish a Jordan-basis framework, including the Jordan chain relation, biorthogonality relations and normalization conventions. Sec.~\ref{sec:puiseux} develops the Puiseux expansion, derives the recursion relations, and obtains the first two eigenvalue corrections in terms of $\p{H}{1}$\,. Sec.~\ref{sec:K1} connects the results to the geometric formalism and expresses the same corrections in terms of $\p{K}{0}_1$\,. Explicit verification for the $N=2$ and $N=3$ cases is carried out in Sec.~\ref{sec:examples}. We summarize our findings and discuss future directions in Sec.~\ref{sec:conc}.

	\section{Setup}
	\label{sec:main}

		Consider a system described by a dimensionless Hamiltonian
		\begin{align}
			\label{PertH}  H(q) = \p{H}{0} + \left(q - q_\mathrm{EP}\right) \p{H}{1} = \p{H}{0} + \epsilon\, \p{H}{1}\,,
		\end{align}
		where $q$ is a continuous parameter, $\p{H}{0} \equiv  H(q_\mathrm{EP})$ is the unperturbed Hamiltonian evaluated at the EP\,, and $\p{H}{1}$ characterizes how the system departs from the EP as $\epsilon = q - q_\mathrm{EP}$ varies. Our primary interest lies in the eigenvalue splitting near an order-$N$ EP. We assume that $\p{H}{0}$ has an eigenvalue $\p{h}{0}$ of algebraic multiplicity $N$ and geometric multiplicity $1$\,, i.e.\ a pure order-$N$ EP. For a generic perturbation, the $N$ eigenvalue branches then split with the same fractional exponent $\epsilon^{1/N}$~\cite{Welters2011}, the precise genericity condition being given in Appendix~\ref{app:generic}. In situations beyond the assumptions made here, including a mixed EP composed of Jordan blocks of different sizes or a failure of the genericity condition, the branches would instead split with different exponents (see, e.g., Refs.~\cite{Moro1997,Seyranian2003,Demange2012,Wang2025,Mandal2026}). Throughout this work, we focus on a pure order-$N$ EP under the generic condition only. At such an EP, the underlying Hamiltonian $\p{H}{0}$ is not diagonalizable; rather, it features a Jordan basis of right generalized eigenvectors $\{\Ket{b_k}\}_{k=0}^{N-1}$ obeying the right Jordan chain relations
		\begin{align}
		 (\p{H}{0}-\p{h}{0})\Ket{b_0} &= 0\,,\\
			(\p{H}{0} - \p{h}{0})\Ket{b_k} &=\Ket{b_{k-1}}\,,
		\end{align}
		for $k = 1, 2, \dots, N-1$. For convenience, we repackage these two equations into
		\begin{align}
			\label{JCR} (\p{H}{0}-\p{h}{0}) \Ket{b_k}=(1-\delta_{k,0})\Ket{b_{k-1}}\,,
		\end{align}
		where $\delta_{k,0}$ is the Kronecker delta. To complement this setup, the left generalized eigenvectors $\{\Bra{b_j}\}_{j=0}^{N-1}$ are defined via the corresponding left Jordan chain relation of $\p{H}{0}$\,,
		\begin{align}
			\label{LJCR} \Bra{b_k}(\p{H}{0}-\p{h}{0}) = (1-\delta_{k,0})\Bra{b_{k-1}}\,,
		\end{align}
		Here, $\Ket{b_0}$ and $\Bra{b_0}$ constitute the right and left eigenvectors at the EP, respectively. With the normalization convention $\Braket{b_0}{b_{N-1}} = 1$\,, the left and right generalized eigenvectors are biorthogonal,
		\begin{align}
			\label{ORTH} \Braket{b_j}{b_k}  = \delta_{j+k,\,N-1}\,,
		\end{align}
		which follows directly from the Jordan chain structures of the left and right bases. In particular, the self-orthogonality $\Braket{b_0}{b_0} = 0$ of the eigenvector reflects that the system is at an EP. Under this setup, the completeness relation takes the form
		\begin{align}
			\label{COMP} \sum_{k=0}^{N-1} \Ket{b_k}\Bra{b_{N-k-1}} = \mathbf{1}\,,
		\end{align}
		replacing the standard spectral decomposition.

		It should be emphasized that our general derivations rely only on the relations~\eqref{JCR}--\eqref{COMP}. Nonetheless, for the concrete computations in the examples below and the specific proofs in Appendix~\ref{app:generic}, it is convenient to adopt an explicit matrix representation. In the Jordan basis itself, the unperturbed Hamiltonian $\p{H}{0}$ takes the Jordan normal form
		\begin{align}
			\label{JordanForm} \p{H}{0} = \begin{pmatrix} \p{h}{0} & 1 & & \\ & \p{h}{0} & \ddots & \\ & & \ddots & 1 \\ & & & \p{h}{0} \end{pmatrix}_{N\times N}\,,
		\end{align}
		and the right and left generalized eigenvectors take the component form
		\begin{align}
			\label{eq:component}\left(\Ket{b_k}\right)_j = \delta_{kj}\,, \qquad \left(\Bra{b_k}\right)_j = \delta_{j,\,N-1-k}\,.
		\end{align}
		The perturbation $\p{H}{1}$ is then treated as a general matrix in this basis. 

		Since $\{\Ket{b_k}\}_{k=0}^{N-1}$ forms a complete basis for the Hilbert space, any physical state $\Ket{\psi}$, regardless of its proximity to the EP, can be expanded as
		\begin{align}
		 \Ket{\psi} = \sum_{k=0}^{N-1} c_k \Ket{b_k}\,,
		\end{align}
		where the coefficients are extracted via biorthogonality,
		\begin{align}
		 c_k = \Braket{b_{N-k-1}}{\psi}\,.
		\end{align}
		Crucially, unlike the usual orthonormal case, the coefficient of $\Ket{b_k}$ is obtained by projecting with $\Bra{b_{N-k-1}}$ rather than $\Bra{b_k}$, as required by the biorthogonality~\eqref{ORTH}.

	\section{Puiseux Expansion of Eigenvalues and Eigenstates}
		\label{sec:puiseux}

		\par In this section, we establish the framework to solve the eigenvalue equation perturbatively,
		\begin{align}
			\label{EGEQ}H(\epsilon)\Ket{h(\epsilon)} = h(\epsilon)\Ket{h(\epsilon)}\,.
		\end{align}
		Here, $H(\epsilon)$ is the perturbed Hamiltonian of Eq.~\eqref{PertH}, and $h(\epsilon)$ and $\Ket{h(\epsilon)}$ are the corresponding perturbed eigenvalue and eigenstate, respectively.

		Since the standard Rayleigh--Schr\"{o}dinger perturbation theory breaks down at an EP due to the incomplete eigenbasis, we instead seek the perturbed eigenvalue and eigenstate via a Puiseux expansion in
		$\epsilon^{1/N}$:
		\begin{align}
		\label{Pert2} h(\epsilon)
		 =\sum_{r=0}^{\infty}\epsilon^{r/N} \p{h}{r/N},~~
		 \Ket{h(\epsilon)}
		 =\sum_{r=0}^{\infty} \epsilon^{r/N} \Ket{\p{p}{r/N}},
		\end{align}
		where $\p{h}{0}$ and $\Ket{\p{p}{0}}\equiv\Ket{b_0}$ correspond to the unperturbed eigenvalue and eigenstate, respectively, and each correction of the eigenstate is expanded in the Jordan basis as
		\begin{align}
			\label{pExp} \Ket{\p{p}{r/N}} = \sum_{k=0}^{N-1} \p{\alpha}{r/N}_k \Ket{b_k}\,,
		\end{align}
		from which, by definition of $\Ket{\p{p}{0}}$\,, it follows that
		\begin{align}
			\label{APL}\p{\alpha}{0}_k=\delta_{0,k}\,.
		\end{align}
		The coefficient $\p{\alpha}{r/N}_0$ appearing in the expansion Eq.~\eqref{pExp} is a free parameter at each order, reflecting the normalization freedom of the perturbed eigenstate. To see this, note that if $\Ket{h(\epsilon)}$ is a solution to Eq.~\eqref{EGEQ}, then so is $\lambda(\epsilon)\Ket{h(\epsilon)}$ for any scalar function $\lambda(\epsilon)$\,. Expanding $\lambda(\epsilon) = 1 + \sum_{r=1}^\infty \p{\lambda}{r/N}\epsilon^{r/N}$ and comparing coefficients at each order $\epsilon^{r/N}$\,, the rescaled eigenstate has expansion coefficients
		\begin{align}
		 \p{\tilde{\alpha}}{r/N}_k = \p{\alpha}{r/N}_k + \sum_{r'=1}^{r}\p{\lambda}{r'/N} \p{\alpha}{(r-r')/N}_k\,,
		\end{align}
		so that $\p{\tilde{\alpha}}{r/N}_k$ is shifted by an additional term involving $\p{\lambda}{r/N}$\,. One can show that, order by order, this provides the gauge freedom to shift $\p{\alpha}{r/N}_0$ to a desired value, reflecting the normalization of the eigenstate. We fix it by the normalization convention $\Braket{b_{N-1}}{h(\epsilon)} = 1$ to all orders in $\epsilon$\,, which sets
		\begin{align}
			\label{NMLZ}\p{\alpha}{r/N}_0=\delta_{0,r}\,.
		\end{align}

		Inserting into the eigenvalue equation Eq.~\eqref{EGEQ}, the left-hand side expands as
		\begin{align}
		 &\hspace{-0.1cm}H(\epsilon)\Ket{h(\epsilon)}
		 = \sum_{r=0}^{N-1}\epsilon^{r/N} \p{H}{0}\Ket{\p{p}{r/N}}\\
		\notag &\hspace{1.7cm}+\sum_{r=N}^{\infty}\hspace{-0.05cm}\epsilon^{r/N} \hspace{-0.1cm}\left(\p{H}{0}\hspace{-0.06cm}\Ket{\p{p}{r/N}}\hspace{-0.06cm}+\p{H}{1}\hspace{-0.06cm}\Ket{\p{p}{r/N-1}}\right)\,,
		\end{align}
		while the right-hand side expands as
		\begin{align}
			h(\epsilon)\Ket{h(\epsilon)}
		 &
			=\sum_{r=0}^{\infty}\epsilon^{r/N} \sum_{r'=0}^{r} \p{h}{r'/N}\Ket{\p{p}{(r-r')/N}}\,.
		\end{align}

		Equating both sides order by order, we can organize the results by order in $\epsilon$ into the following three cases:
		\begin{itemize}
			\item $\underline{\epsilon^{r/N}~\mbox{with}~r=0}:$
				\begin{align}
				 \left(\p{H}{0}-\p{h}{0}\right)\Ket{\p{p}{0}}=0\,.
				\end{align}
		 \item $\underline{\epsilon^{r/N}~\mbox{with}~0<r<N}:$
				\begin{align}
					\label{RC1} (\p{H}{0}-\p{h}{0})\Ket{\p{p}{r/N}}=\sum_{r'=1}^{r} \p{h}{r'/N} \Ket{\p{p}{(r-r')/N}}\,.
				\end{align}

			\item $\underline{\epsilon^{r/N}~\mbox{with}~r\geq N}:$
				\begin{align}
					&\label{RC2} \hspace{-0.55cm}(\p{H}{0}-\p{h}{0})\Ket{\p{p}{r/N}}\\
					&\hspace{-0.4cm}=-\p{H}{1}\Ket{\p{p}{r/N-1}}
					\notag +\sum_{r'=1}^{r} \p{h}{r'/N} \Ket{\p{p}{(r-r')/N}}\,.
				\end{align}
		\end{itemize}
		Since the $r = 0$ case is trivially satisfied by the unperturbed state, we restrict our attention to $r > 0$ in the remainder of this section. We aim to express $\p{h}{r/N}$ in terms of $\p{h}{0}$ and the matrix elements of $\p{H}{1}$\,. To extract this information, we apply $\Bra{b_j}$ to Eqs.~\eqref{RC1} and \eqref{RC2}. Employing the Jordan chain relation in Eq.~\eqref{JCR} and the biorthogonality in Eq.~\eqref{ORTH}, the left-hand sides of these equations become
		\begin{align}
		 \Bra{b_j}(\p{H}{0} - \p{h}{0})\Ket{\p{p}{r/N}}
		 = \p{\alpha}{r/N}_{N-j}(1-\delta_{0,j})\,.
		\end{align} The right-hand side of Eq.~\eqref{RC1} gives
		\begin{align}
		 \sum_{r'=1}^{r} \p{h}{r'/N}\Braket{b_j}{\p{p}{(r-r')/N}}
		 = \sum_{r'=1}^{r} \p{h}{r'/N}\, \p{\alpha}{(r - r')/N}_{N - j - 1}\,,
		\end{align}
		and, similarly, the right hand side of Eq.~\eqref{RC2}\,,
		\begin{align}
			\notag &-\Bra{b_j}\p{H}{1}\Ket{\p{p}{r/N-1}}+\sum_{r'=1}^{r}\p{h}{r'/N}\Braket{b_j}{\p{p}{(r-r')/N}}\\
			& = - \sum_{k=0}^{N-1}[[\p{H}{1}]]_{jk}\p{\alpha}{r/N - 1}_k+\sum_{r'=1}^{r}\p{h}{r'/N}\, \p{\alpha}{(r - r')/N}_{N - j - 1}\,,
		\end{align}
		where $[[\p{H}{1}]]_{ij}$ is the short hand notation for the matrix element $\Bra{b_i}\p{H}{1}\Ket{b_j}$ of $\p{H}{1}$ in the Jordan basis.

		Equating left and right of Eqs.~\eqref{RC1} and \eqref{RC2}, we obtain the recursion relations:
		\begin{itemize}
		 \item $\underline{0<r<N}$\\
				\begin{align}
				 \label{NRC1} \p{\alpha}{r/N}_{N-j}(1-\delta_{0,j})=\sum_{r'=1}^{r} \p{h}{r'/N}\, \p{\alpha}{(r - r')/N}_{N-j-1}\,,
				\end{align}
		 \item $\underline{r\geq N}$\\
				\begin{align}
					\label{NRC2} \p{\alpha}{r/N}_{N-j}(1-\delta_{0,j})&= - \sum_{k=0}^{N-1}[[\p{H}{1}]]_{jk} \p{\alpha}{r/N - 1}_k\\
				\notag & \quad + \sum_{r'=1}^{r} \p{h}{r'/N}\, \p{\alpha}{(r-r')/N}_{N-j-1}\,.
				\end{align}
		\end{itemize}

		The eigenvalue corrections $\p{h}{r/N}$ can be obtained iteratively from these two recursion relations. To illustrate this procedure, we take the base case $\p{h}{1/N}$ as a concrete starting point. The key step is to establish the explicit connection between $\p{\alpha}{r/N}_k$ and $\p{h}{r/N}$\,. First, by considering Eq.~\eqref{NRC1} with $j = N-1$ and applying the normalization in Eq.~\eqref{NMLZ}, only the $r' = r$ term survives on the right-hand side, yielding
		\begin{align}
		 \p{\alpha}{r/N}_1 = \p{h}{r/N}\,,\quad(0< r< N)\,.
		 \label{eq:alpha1}
		\end{align}

		Similarly, by taking $j = N-2$ in Eq.~\eqref{NRC1}, one finds
		\begin{align}
		 \p{\alpha}{r/N}_2 &\hspace{-1mm} = \hspace{-1mm}\sum_{r'=1}^{r} \p{h}{r'/N}\,\p{\alpha}{(r - r')/N}_1 \hspace{-1mm}=\hspace{-1mm}\sum_{r'=1}^{r-1} \p{h}{r'/N}\,\p{h}{(r-r')/N}\,,
		\end{align}
		where in the last equality we have used Eq.~\eqref{APL} and substituted Eq.~\eqref{eq:alpha1}\,. In other words, it is now clear that
		\begin{align}
		 \p{\alpha}{1/N}_2 &=0\,,\\
			\notag \p{\alpha}{r/N}_2 & = \sum_{r'=1}^{r-1} \p{h}{r'/N}\,\p{h}{(r-r')/N}\,,\quad (1< r<N)\,.
		\end{align}

		Iterating the procedure for $j=N-3$\,, one obtains,
		\begin{align}
			\p{\alpha}{1/N}_3 & = \p{\alpha}{2/N}_3 = 0\,,\\[2mm]
			\notag \p{\alpha}{r/N}_3 &
		     =\hspace{-0.5cm}\sum_{r_1+r_2+r_3=r\atop{r_1,r_2,r_3\geq 1}}\hspace{-0.5cm}\p{h}{r_1/N}\p{h}{r_2/N}\p{h}{r_3/N}\,,\quad (2< r< N)\,,
		\end{align}
		with the sum running over all partitions of $r$ into three positive integers. Given the examples shown above, the pattern is clear: iterating this procedure for $j = N-4, N-5, \ldots, 1$ successively, we find that\,,
		\begin{align}
			\notag \p{\alpha}{r/N}_m&=0\,,  \hspace{3.65cm}(0<r < m<N) \,,\\
			\notag \p{\alpha}{r/N}_m &= \hspace{-0.35cm}\sum_{\substack{r_1+\cdots+r_m=r\\r_i\geq 1}}\hspace{-0.35cm} \p{h}{r_1/N}\cdots \p{h}{r_m/N}\,,\,\, (0<m \leq r<N)\,,\\[-0.5cm]
			\label{iter} 
		\end{align}
		which is a sum over all partitions of $r$ into $m$ positive integer parts. Now we have expressed all $\p{\alpha}{r/N}_k$ for $0<r<N$ in terms of $\p{h}{r/N}$'s. In particular, when $r = m$\,, the unique partition is $r_1 = \cdots = r_m = 1$\,, giving
		\begin{align}
		\label{alphamm}
			\p{\alpha}{m/N}_m = \left(\p{h}{1/N}\right)^m, \hspace{1cm} (0< m<N)\,.
		\end{align}
		
		Having the information of $\p{\alpha}{r/N}_k$\,, we can obtain $\p{h}{r/N}$ order by order iteratively by Eq.~\eqref{NRC2}. To begin, the leading order $\p{h}{1/N}$ can be determined by taking $j = 0$ and $r = N$ in Eq.~\eqref{NRC2},
		\begin{align}
		\label{h1N} 0 = -\sum_{k=0}^{N-1}[[\p{H}{1}]]_{0k}\,\p{\alpha}{0}_k
		 + \sum_{r'=1}^{N}\p{h}{r'/N}\,\p{\alpha}{(N-r')/N}_{N-1}\,.
		\end{align}
		Using Eqs.~\eqref{APL} and \eqref{iter}, only $k=0$ survives in the first term and $r'=1$ in the second, respectively, so Eq.~\eqref{h1N} reduces to
		\begin{align}
		 -[[\p{H}{1}]]_{00}+ \left(\p{h}{1/N}\right)^N=0\,,
		\end{align}
		which implies
		\begin{align}
			\notag &\boxed{\,
		 \p{h}{1/N} = \Big|[[\p{H}{1}]]_{00}\Big|^{1/N}
		 \exp\left[i\,\frac{\arg\left([[\p{H}{1}]]_{00}\right) + 2\pi k}{N}\right]
		 }\,,\\[1.5mm]
		 \label{h1NHp} & \hspace{2.5cm}(k = 0, 1, \ldots, N-1)\,.
		\end{align}
		These $N$ values of $\p{h}{1/N}$ correspond to the $N$ branches of the Puiseux expansion, each related by a rotation of $2\pi/N$ in the complex plane. This leading coefficient agrees with the classical perturbation theory of nonsemisimple eigenvalues~\cite{Lidskii1966,VishikLyusternik1960,Moro1997,Seyranian2003}. Importantly, Equation~\eqref{h1NHp} holds under the genericity condition $[[\p{H}{1}]]_{00}\neq0$\,; when it fails, the leading Puiseux exponent is no longer $1/N$ and a separate analysis is needed (see Appendix~\ref{app:generic}).

		Next we consider $\p{h}{2/N}$ by taking $j = 0$ and $r = N+1$ in Eq.~\eqref{NRC2}, which gives
		\begin{align}
		 0 = -\sum_{k=0}^{N-1}[[\p{H}{1}]]_{0k}\, \p{\alpha}{1/N}_k + \sum_{r'=1}^{N+1} \p{h}{r'/N}\,\p{\alpha}{(N+1-r')/N}_{N-1}\,.
		\end{align}
		Non-zero $\p{\alpha}{(N+1-r')/N}_{N-1}$ requires $r'=1,2$\,, allowing the equation to be further simplified via Eq.~\eqref{iter}, which leads to
		\begin{align}
			\label{preh2N} 0 = -[[\p{H}{1}]]_{01}\, \p{\alpha}{1/N}_1 + \p{h}{1/N}\,\p{\alpha}{N/N}_{N-1}+ \p{h}{2/N}\,\p{\alpha}{(N-1)/N}_{N-1}\,.
		\end{align}
		By virtue of Eq.~\eqref{alphamm}, it is straightforward to see that $\p{\alpha}{1/N}_1=\p{h}{1/N}$ and $\p{\alpha}{(N-1)/N}_{N-1}= \left(\p{h}{1/N}\right)^{N-1}$\,. The coefficient $\p{\alpha}{N/N}_{N-1}$\,, however, requires more care, since it lies in the regime $r \geq N$ where $\p{H}{1}$ enters the recursion. Setting $j = 1$ and $r = N$ in the recursion Eq.~\eqref{NRC2},
		\begin{align}
		 \p{\alpha}{N/N}_{N-1} = -[[\p{H}{1}]]_{10} + \sum_{r'=1}^{N}\p{h}{r'/N}\,\p{\alpha}{(N-r')/N}_{N-2}\,.
		\end{align}
		Using Eq.~\eqref{iter}, $\p{\alpha}{(N-r')/N}_{N-2}$ is evaluated by partitioning $N - r'$ into $N-2$ positive integer parts, which is non-zero only for $r' = 1, 2$\,. The two surviving terms give
		\begin{align}
		 \sum_{r'=1}^{N}\p{h}{r'/N}\,\p{\alpha}{(N-r')/N}_{N-2}
		 &= (N-1)\,\p{h}{2/N}\,\left(\p{h}{1/N}\right)^{N-2}\,,
		\end{align}
		hence
		\begin{align}
		 \p{\alpha}{N/N}_{N-1} = -[[\p{H}{1}]]_{10} + (N-1)\,\p{h}{2/N}\,\left(\p{h}{1/N}\right)^{N-2}\,.
		 \label{eq:alphaNN}
		\end{align}
		With these, Eq.~\eqref{preh2N} becomes
		\begin{align}
		 \notag  \hspace{-0.4cm} 0
		 = \big([[\p{H}{1}]]_{01} \hspace{-0.6mm}+ \hspace{-0.6mm}[[\p{H}{1}]]_{10}\big)\p{h}{1/N}
		 \hspace{-0.4mm}-\hspace{-0.4mm} N\p{h}{2/N}\left(\p{h}{1/N}\right)^{N-1}\hspace{-4mm}\,,\\[0.5mm]
		\end{align}
		so that we conclude that
		\begin{align}
		 \boxed{
		 ~ \p{h}{2/N} = \frac{[[\p{H}{1}]]_{01} + [[\p{H}{1}]]_{10}}
		 {N\,\left(\p{h}{1/N}\right)^{N-2}}
		 ~ }\,.
		 \label{eq:h2N_final}
		\end{align}
		This result agrees with the one obtained in~\cite{Welters2011}. The same iterative procedure yields the higher-order coefficients $\p{h}{r/N}$\,. In what follows, we explore the connection between these coefficients and the geometric formalism developed in~\cite{original_paper}.

	\section{Evolution Generator $K_1$ at the Exceptional Point}
	\label{sec:K1}

		Having established the Puiseux expansion of the eigenvalue and eigenstate, we now turn to the $q$-evolution equation~\cite{Ju2024} and study the structure of the eigenvalue-evolution generator $K_1$ near the EP. The key equations from the geometric framework are as follows. The generators $K_0$ and $K_1$ capture the evolution of the eigenstate and the eigenvalue, respectively:
		\begin{align}
		 K_0\Ket{h(\epsilon)} = i\partial_q\Ket{h(\epsilon)}\,,\,
		 K_1\Ket{h(\epsilon)} = (\partial_q h)\Ket{h(\epsilon)}\,.
		 \label{Kstate}
		\end{align}
		They satisfy the relations
		\begin{align}
			[K_1, H] = 0\, ,\qquad K_1 = i[K_0, H] + \partial_q H\,.
		 \label{Kcond}
		\end{align}
		Since $\epsilon = q - q_\mathrm{EP}$\,, we have $\partial_q = \partial_\epsilon$\,. Hereafter, only $\partial_\epsilon$ is used.
		The generator $K_0$ enters Eq.~\eqref{Kcond} only through the commutator $[K_0, H]$, and is therefore defined only up to a residual gauge freedom of the underlying framework~\cite{Ju2024,Ju2025}. As the eigenvalue corrections are gauge-invariant, our results are independent of $K_0$.

		From the Puiseux expansion of $h(\epsilon)$ and $\Ket{h(\epsilon)}$ in Eq.~\eqref{Pert2}, the derivative $\partial_\epsilon\Ket{h(\epsilon)}$ diverges as $\epsilon^{1/N-1}$ when $\epsilon\to0$; by Eq.~\eqref{Kstate}, $K_0$ and $K_1$ must then contain simple poles. While they could in general involve fractional powers of $\epsilon$\,, we find it sufficient to expand them as Laurent series in integer powers of $\epsilon$\,, starting from $\epsilon^{-1}$,
		\begin{align}
		 K_0(\epsilon) = \sum_{m=-1}^{\infty}\epsilon^{m}\p{K}{m}_0, \quad K_1(\epsilon) = \sum_{m=-1}^{\infty}\epsilon^{m}\p{K}{m}_1\,.
		 \label{Kexpand}
		\end{align}
		This expansion is compatible with Eqs.~\eqref{Kcond} and \eqref{Kstate} order by order in $\epsilon^{1/N}$\,. Substituting Eqs.~\eqref{PertH} and \eqref{Kexpand} into Eq.~\eqref{Kcond} and collecting terms at order $\epsilon^{m}$:

		\begin{itemize}
		\item \underline{$m = -1$}:
		\begin{align}
		 [\p{K}{-1}_1,\, \p{H}{0}] = 0\,,
		 \label{commneg}
		\end{align}
		\begin{align}
		 \p{K}{-1}_1 = i\left[\p{K}{-1}_0,\, \p{H}{0}\right]\,.
		 \label{K1expneg}
		\end{align}

		\item \underline{$m \geq 0$}:
		\begin{align}
		 [\p{K}{m}_1,\, \p{H}{0}] + [\p{K}{m-1}_1,\, \p{H}{1}] = 0\,,
		 \label{commpos}
		\end{align}
		\begin{align}
		 \p{K}{m}_1 = i\left[\p{K}{m}_0,\, \p{H}{0}\right]
		 + i\left[\p{K}{m-1}_0,\, \p{H}{1}\right]+\p{H}{1}\delta_{0,m}\,.
		 \label{K1exppos}
		\end{align}

		\end{itemize}

		In the same manner, matching powers of $\epsilon^{1/N}$ on both sides of Eq.~\eqref{Kstate}, we have
		\begin{flalign}
		 &\sum_{\mathcal{V}_s} 
		 \p{K}{m}_1\Ket{\p{p}{r'/N}}
		 =\sum_{\mathcal{T}_s}
		  \frac{r}{N}\,\p{h}{r/N}\Ket{\p{p}{r'/N}}\,,
		 \label{eq:K1n_expand}
		 \\[0.1cm]
		 &\sum_{\mathcal{V}_s}
		 \p{K}{m}_0\Ket{\p{p}{r'/N}}
		 = i\,\frac{s+N}{N}\Ket{\p{p}{(s+N)/N}}\,,
		 \label{eq:K0n_expand}
		\end{flalign}
		where $\mathcal{V}_s = \{(m,r') : Nm+r'=s,\ m\geq -1,\ r'\geq 0\}$ and $\mathcal{T}_s = \{(r,r') : r+r'=s+N,\ r\geq 1,\ r'\geq 0\}$\,. The right-hand side of Eq.~\eqref{eq:K1n_expand} vanishes when $s = -N$\,, since $\mathcal{T}_{-N}$ is empty ($r \geq 1$ cannot be satisfied when $r + r' = 0$)\,. In particular, $\mathcal{V}_{-N}$ then contains only the single term $(m,r') = (-1,0)$\,, giving
		\begin{align}
		 \p{K}{-1}_0\Ket{b_0} = 0\,,
		 \qquad
		 \p{K}{-1}_1\Ket{b_0} = 0\,.
		 \label{eq:K_leading}
		\end{align}
		It is useful to write down the projection of Eq.~\eqref{eq:K1n_expand} onto $\Bra{b_j}$ at order $\epsilon^{s/N}$; using biorthogonality Eq.~\eqref{ORTH}, we obtain
		\begin{align}
			\sum_{\mathcal{V}_s}
		 \sum_{k=0}^{N-1}
		 \p{\alpha}{r'/N}_k\,[[\p{K}{m}_1]]_{jk}
		 =
		 \sum_{\mathcal{T}_s}
		 \frac{r}{N}\,\p{h}{r/N}\,\p{\alpha}{r'/N}_{N-1-j}\,.
		 \label{eq:K1proj}
		\end{align}
		We now use these equations to extract the eigenvalue corrections $\p{h}{r/N}$ in terms of $\p{K}{m}_1$ order by order. First, to obtain $\p{h}{1/N}$\,, we take Eq.~\eqref{eq:K1proj} with $(s,j)=(0,0)$\,. The left-hand side receives contributions only from $(m,r') = (0,0)$ and $(m,r') = (-1,N)$\,, so that,
		\begin{align}
		 [[\p{K}{0}_1]]_{00}
		 + \p{\alpha}{N/N}_{N-1}\,[[\p{K}{-1}_1]]_{0,N-1}
		 =
		 \sum_{\mathcal{T}_0}
		 \frac{r}{N}\,\p{h}{r/N}\,\p{\alpha}{r'/N}_{N-1}\,.
		 \label{s0_proj}
		\end{align}
		The second term on the left-hand side drops out by Eq.~\eqref{Km1}, so that Eq.~\eqref{s0_proj} reduces to
		\begin{align}
		 [[\p{K}{0}_1]]_{00} 
		 =\sum_{N\geq r\geq 1}
		 \frac{r}{N}\,\p{h}{r/N}\,\p{\alpha}{(N-r)/N}_{N-1}\,.
		 \label{eq:s0_reduced}
		\end{align}
		Note that $[[\p{K}{0}_1]]_{00}$ itself does not vanish, in contrast to the $[[\p{K}{-1}_1]]_{00}$ term, which vanishes as shown in Appendix~\ref{app:Km1}. This is because $\p{K}{0}_1$ satisfies the relation $[\p{K}{0}_1, \p{H}{0}] + [\p{K}{-1}_1, \p{H}{1}] = 0$ rather than $[\p{K}{0}_1, \p{H}{0}] = 0$\,. To simplify the right-hand side, recall that $\p{\alpha}{(N-r)/N}_{N-1}$ is non-zero only when $r \leq 1$ according to Eq.~\eqref{iter}. Together with the constraint $r \geq 1$ from the summation, only $r = 1$ contributes, leading to
		\begin{align}
			[[\p{K}{0}_1]]_{00}
			= \frac{1}{N}\,\p{h}{1/N}\,\p{\alpha}{(N-1)/N}_{N-1}
			= \frac{\left(\p{h}{1/N}\right)^N}{N}\,,
		\end{align}
		where the last equality follows from Eq.~\eqref{alphamm}. Consequently,
		\begin{align}
			\boxed{
				~ \left(\p{h}{1/N}\right)^N = N\,[[\p{K}{0}_1]]_{00}
			~}\,.
			\label{eq:h1N_K1}
		\end{align}
		This is the EP analogue of the non-EP result in~\cite{original_paper}: whereas in the non-EP case the first-order eigenvalue correction is given by the diagonal matrix element of $\p{K}{0}_1$\,, near the EP the $N$-th power of the leading correction is $N$ times the lower-triangular matrix element of $\p{K}{0}_1$\,.

		\medskip

		The same projection technique yields the second Puiseux coefficient $\p{h}{2/N}$ in terms of $K_1$\,. Consider Eq.~\eqref{eq:K1proj} with\\
		~~\\
		$\underline{(s,j)=(1,0)}:$
		\begin{align}
			\notag&\sum_{k=0}^{N-1} \p{\alpha}{1/N}_k[[\p{K}{0}_1]]_{0k}
			+ \sum_{k=0}^{N-1} \p{\alpha}{(N+1)/N}_k[[\p{K}{-1}_1]]_{0k}\\
			 & \quad = \sum_{\mathcal{T}_1} \frac{r}{N}\,\p{h}{r/N}\,\p{\alpha}{r'/N}_{N-1}\,,
		\end{align}
		$\underline{(s,j)=(0,1)}:$
		\begin{align}
			\notag&\sum_{k=0}^{N-1} \p{\alpha}{0}_k\,[[\p{K}{0}_1]]_{1k}
			+ \sum_{k=0}^{N-1} \p{\alpha}{N/N}_k[[\p{K}{-1}_1]]_{1k}\\
			& \quad = \sum_{\mathcal{T}_0} \frac{r}{N}\,\p{h}{r/N}\,\p{\alpha}{r'/N}_{N-2}\,.
		\end{align}
		For $(s,j) = (1,0)$\,, the left-hand side receives contributions from both $(m,r') = (0,1)$ and $(m,r') = (-1,N+1)$\,. However, the second term on the left-hand side vanishes according to Eq.~\eqref{Km1}. On the right-hand side, the condition $\p{\alpha}{r'/N}_{N-1} \neq 0$ demands $r' \geq N-1$\,, which implies $r \leq 2$\,. Collectively, these conditions lead to
		\begin{align}
		 [[\p{K}{0}_1]]_{01}
		 = \frac2N\,\p{h}{2/N}\,\left(\p{h}{1/N}\right)^{N-2}
		 + \frac1N\,\p{\alpha}{N/N}_{N-1}\,.
		 \label{eq:proj10}
		\end{align}
		Turning to $(s,j) = (0,1)$\,, the left-hand side retains only the terms $(m,r') = (0,0)$ and $(m,r') = (-1,N)$\,. Applying Eqs.~\eqref{Km1} and \eqref{eq:K1m1_super} then allows these terms to be further simplified, yielding
		\begin{align}
			\begin{split}
			 [[\p{K}{0}_1]]_{10} & = \frac2N\,\p{h}{2/N}\,\left(\p{h}{1/N}\right)^{N-2}\\
			 & \quad + \frac1N\,\p{h}{1/N}\,\p{\alpha}{(N-1)/N}_{N-2} - \frac1N\,\p{\alpha}{N/N}_{N-1}\,.
			 \end{split}
		 \label{eq:proj01}
		\end{align}
		Adding Eqs.~\eqref{eq:proj10} and \eqref{eq:proj01}, the unknown $\p{\alpha}{N/N}_{N-1}$ cancels. The remaining coefficient is fixed by Eq.~\eqref{iter}: partitioning $N-1$ into $N-2$ positive parts gives $\p{\alpha}{(N-1)/N}_{N-2} = (N-2)\,\p{h}{2/N}\,\left(\p{h}{1/N}\right)^{N-3}$\,, so that
		\begin{align}
		 [[\p{K}{0}_1]]_{01} + [[\p{K}{0}_1]]_{10}
		 = \frac{N+2}{N}\,\p{h}{2/N}\,\left(\p{h}{1/N}\right)^{N-2}\,.
		\end{align}
		We thus obtain $\p{h}{2/N}$ entirely in terms of the evolution generator,
		\begin{align}
		 \boxed{
		 \p{h}{2/N}
		 = \frac{N}{N+2}\,
		 \frac{[[\p{K}{0}_1]]_{01} + [[\p{K}{0}_1]]_{10}}{\left(\p{h}{1/N}\right)^{N-2}}
		 }\,.
		 \label{eq:h2N_K1}
		\end{align}
		Equation~\eqref{eq:h2N_K1} is the $K_1$ counterpart of the $\p{H}{1}$ expression in Eq.~\eqref{eq:h2N_final}, with the two being related through the identities in Eq.~\eqref{commpos} that connect $[[\p{K}{0}_1]]$ to $[[\p{H}{1}]]$\,. One can proceed in a similar manner to obtain higher-order eigenvalue perturbations.

	\section{Examples}
	\label{sec:examples}

		We verify the general results of Secs.~\ref{sec:puiseux} and \ref{sec:K1} by working out the $N = 2$ and $N = 3$ cases explicitly. In each case we confirm both the $\p{H}{1}$ representations [Eqs.~\eqref{h1NHp} and \eqref{eq:h2N_final}] and the $\p{K}{0}_1$ representations [Eqs.~\eqref{eq:h1N_K1} and \eqref{eq:h2N_K1}].

		\subsection{The $N = 2$ Case}

			As an $N=2$ toy model, we start with the unperturbed Hamiltonian $\p{H}{0}$ in Jordan normal form and the perturbation $\p{H}{1}$ in the same basis,
			\begin{align}
			 \p{H}{0} \doteq \begin{pmatrix} \p{h}{0} & 1 \\ 0 & \p{h}{0} \end{pmatrix}\,,
			 \quad
			 \p{H}{1} \doteq \begin{pmatrix} \mu_1 & \mu_2 \\ \mu_3 & \mu_4 \end{pmatrix}\,,
			\end{align}
			where, by Eq.~\eqref{eq:component}, the basis vectors are $\Ket{b_0}=(1,0)$, $\Ket{b_1}=(0,1)$, $\Bra{b_0} = (0,1)$, and $\Bra{b_1} = (1,0)$\,. One can directly verify that the leading- and sub-leading-order correction formulas, Eqs.~\eqref{h1NHp} and \eqref{eq:h2N_final}, yield
			\begin{align}
			 \left(\p{h}{1/2}\right)^2 = \mu_3 \,, \quad \p{h}{1} = \frac{\mu_1+\mu_4}{2} \,.
			\end{align}
			This is consistent with the Puiseux expansion of the exact eigenvalues. Next, we check the validity of Eqs.~\eqref{eq:h1N_K1} and \eqref{eq:h2N_K1}, which involves computing specific entries of $\p{K}{0}_1$\,. This can be achieved by parameterizing $\p{K}{0}_1$ as
			\begin{align}
			 \p{K}{0}_1 = \begin{pmatrix} \nu_1 & \nu_2 \\ \nu_3 & \nu_4 \end{pmatrix} \,,
			\end{align}
			and similarly parameterizing the matrices $\p{K}{-1}_1$\,, $\p{K}{-1}_0$\,, and $\p{K}{0}_0$\,. Inserting these expressions into Eqs.~\eqref{commneg} and \eqref{K1expneg}, together with Eqs.~\eqref{commpos} and \eqref{K1exppos} at $m=1$\,, we arrive at
			\begin{align}
			 \p{K}{0}_1 = \begin{pmatrix} \displaystyle\frac{3\mu_1+\mu_4}{4} & \nu_2 \\ \displaystyle\frac{\mu_3}{2} & \displaystyle\frac{\mu_1+3\mu_4}{4} \end{pmatrix} \,.
			\end{align}
			Applying these results to Eqs.~\eqref{eq:h1N_K1} and \eqref{eq:h2N_K1} with $N=2$\,, we find that the results are fully consistent.

		\subsection{The $N = 3$ Case}

			Similarly, for $N = 3$\,, we consider the unperturbed Hamiltonian $\p{H}{0}$ and the perturbation $\p{H}{1}$\,,
			\begin{align}
			 \p{H}{0} \doteq \begin{pmatrix} \p{h}{0} & 1 & 0 \\ 0 & \p{h}{0} & 1 \\ 0 & 0 & \p{h}{0} \end{pmatrix},
			 \qquad
			 \p{H}{1} \doteq \begin{pmatrix}
			 \kappa_1 & \kappa_2 & \kappa_3 \\
			 \kappa_4 & \kappa_5 & \kappa_6 \\
			 \kappa_7 & \kappa_8 & \kappa_9
			 \end{pmatrix}\,,
			 \label{eq:H3x3}
			\end{align}
			The basis vectors can be read off from Eq.~\eqref{eq:component} as $\Ket{b_0} = (1,0,0)$\,, $\Ket{b_1} = (0,1,0)$\,, $\Ket{b_2} = (0,0,1)$\,, $\Bra{b_0} = (0,0,1)$\,, $\Bra{b_1} = (0,1,0)$\,, and $\Bra{b_2} = (1,0,0)$\,. Following the same procedure as before, one can check that the leading- and sub-leading-order correction formulas, Eqs.~\eqref{h1NHp} and \eqref{eq:h2N_final} for $N=3$\,, yield
			\begin{align}
				\left(\p{h}{1/3}\right)^3 = \kappa_7 \,, \quad \p{h}{2/3} = \frac{\kappa_4+\kappa_8}{3\p{h}{1/3}} \,.
			\end{align}
			To verify Eqs.~\eqref{eq:h1N_K1} and \eqref{eq:h2N_K1}, we evaluate the specific entries of $\p{K}{0}_1$ by adopting the parameterization
			\begin{align}
			 \p{K}{0}_1 = \begin{pmatrix} \omega_1 & \omega_2 & \omega_3 \\ \omega_4 & \omega_5 & \omega_6 \\ \omega_7 & \omega_8 & \omega_9 \end{pmatrix} \,,
			\end{align}
			along with similar parameterizations for $\p{K}{-1}_1$\,, $\p{K}{-1}_0$\,, and $\p{K}{0}_0$\,. Feeding these matrices into Eqs.~(\ref{commneg}-\ref{K1exppos}) at $m=1$\,, we obtain
			\begin{align}
			 \notag
			 \p{K}{0}_1 \doteq \begin{pmatrix}
			 \omega_1 & \omega_2 & \omega_3 \\[10pt]
			 \dfrac{4\kappa_4+\kappa_8}{9} & \omega_5 & \omega_2+\omega_{10} \\[10pt]
			 \dfrac{\kappa_7}{3} & \dfrac{\kappa_4+4\kappa_8}{9} & \omega_9
			 \end{pmatrix} \,,
			\end{align}
			with the diagonal entries determined as
			\begin{align}
			 \notag \omega_1 &= \dfrac{\kappa_4^2 + 15\kappa_1\kappa_7 + 6\kappa_5\kappa_7 - \kappa_4\kappa_8 - 2\kappa_8^2 + 6\kappa_7\kappa_9}{27\kappa_7} \,, \\
			 \notag \omega_5 &= \dfrac{\kappa_4^2 + 6\kappa_1\kappa_7 + 15\kappa_5\kappa_7 + 2\kappa_4\kappa_8 + \kappa_8^2 + 6\kappa_7\kappa_9}{27\kappa_7} \,, \\
			 \notag \omega_9 &= \dfrac{-2\kappa_4^2 + 6\kappa_1\kappa_7 + 6\kappa_5\kappa_7 - \kappa_4\kappa_8 + \kappa_8^2 + 15\kappa_7\kappa_9}{27\kappa_7} \,,\\
			 \omega_{10}&=-\dfrac{(\kappa_4+\kappa_8)(\kappa_1-\kappa_9) + 3(\kappa_2-\kappa_6)\kappa_7}{9\kappa_7} 
			 \,.
			\end{align}
			Inserting these results into Eqs.~\eqref{eq:h1N_K1} and \eqref{eq:h2N_K1} with $N=3$ confirms that the consistency still holds.

	\section{Conclusion}
	\label{sec:conc}

		We have developed a systematic Rayleigh--Schr\"{o}dinger-like perturbation theory for non-Hermitian quantum systems at an EP of order $N$\,. Working in the Jordan basis of the unperturbed Hamiltonian $\p{H}{0}$ and expanding the perturbed eigenvalue and eigenstate in Puiseux series in $\epsilon^{1/N}$\,, we derived explicit recursion relations for the expansion coefficients and solved them iteratively. The first two eigenvalue corrections are obtained in two complementary forms. In terms of the perturbation $\p{H}{1}$\,, they are given by Eqs.~\eqref{h1NHp} and \eqref{eq:h2N_final}; the expression for $\p{h}{2/N}$ is consistent with the result of~\cite{Welters2011}, providing an independent check of our results. In terms of the evolution generator $K_1$ of the geometric formalism~\cite{original_paper,Ju2024}, the same coefficients are given by Eqs.~\eqref{eq:h1N_K1} and \eqref{eq:h2N_K1}, which together constitute the EP analogue of the non-EP result of~\cite{original_paper}. Both representations were verified explicitly for the $N = 2$ and $N = 3$ cases.

		Several directions for future investigation are suggested by our results. The recursion relations derived here provide, in principle, a complete algorithm for computing the Puiseux expansion to arbitrarily high order; expressing the general coefficient $\p{h}{r/N}$ entirely in terms of $K_1$ matrix elements would yield a compact formula at all orders. A related and important direction concerns cases where the leading $\epsilon^{1/N}$ scaling assumed throughout breaks down. This scaling relies on the generic condition $[[\p{H}{1}]]_{00} \neq 0$ derived in Appendix~\ref{app:generic}; when it fails, the leading Puiseux exponent changes. For instance, an order-$3$ block can produce an $\epsilon^{1/2}$ splitting. This direction has long been analyzed using Newton-diagram techniques~\cite{Moro1997,Seyranian2003}, and has been continued in studies relating the leading exponent to the structure of $\p{H}{1}$ in the Jordan basis~\cite{Demange2012,Wang2025,Mandal2026,Znojil2026}. While our results only establish the equivalence of this genericity condition with $[[\p{H}{1}]]_{00} \neq 0$, a natural next step is to extend the framework to the nongeneric regime, where the leading exponent differs.

		Beyond these mathematical extensions, the Jordan-chain formalism may be relevant to the computation of entanglement entropy directly at an EP. Recent work on composite non-Hermitian systems has shown that higher-order EPs can host entangled eigenstates and induce disentanglement during time evolution~\cite{WiersigChen2025}. These analyses, however, do not build on the Jordan-chain structure that underlies the defective EP. Even away from EPs, entanglement in non-Hermitian systems is already subtle owing to the biorthogonal structure of the eigenbasis~\cite{Bianchini2015,Couvreur2017,Tu2022}. This is compounded at an EP, where the Hamiltonian becomes defective: a faithful description then calls for the full Jordan chain of generalized eigenvectors. The Jordan-chain structure has been exploited to characterize other exceptional-point quantities, such as the spectral response strength and the Petermann factor~\cite{WiersigResponse2022,KulligWiersigSchomerus2025}. Nevertheless, a systematic treatment of entanglement based on this framework remains largely unexplored, a direction to which our formalism could be extended. In a different direction, the Jordan-basis structure introduced here may suggest a natural entry point for extending the quantum mechanics bootstrap program~\cite{Han:2020bkb,Berenstein:2021loy,Li:2022prn,Khan:2024mhc} to the exceptional-point setting, where eigenstates are replaced by the Jordan chain of generalized eigenvectors and the bootstrap constraints must be reformulated accordingly.

		Finally, the present framework may offer an alternative understanding of broader physical phenomena, such as the nontrivial phase accumulated when encircling an EP~\cite{Dembowski2004,Arkhipov2023,Lai2024} and the topological structure of the eigenvalue Riemann surface~\cite{Heiss2012,Bergholtz2021,Ju2025a}. We hope that it will serve as a useful starting point for these and other investigations of perturbation theory at EPs.

	\begin{acknowledgments}
		This work was supported by the National Science and Technology Council (NSTC). W.M.C. received partial support through Grant No. NSTC 113-2811-M-110-008; C.Y.J. received partial support through Grant No. NSTC 114-2112-M-029-003; and both authors were partially supported by Grant No. NSTC 112-2112-M-110-013-MY3.

	\end{acknowledgments}

	\begin{appendix}
		\section{Structure of $\p{K}{-1}_1$}
			\label{app:Km1}
			We derive the structure of $\p{K}{-1}_1$\,, which is needed for the derivation of the perturbed eigenvalue in the main text. To begin, projecting Eq.~\eqref{commneg} between $\Bra{b_i}$ and $\Ket{b_j}$ and applying the Jordan-chain relation Eq.~\eqref{JCR} on both the ket and the bra (the $\p{h}{0}$ terms cancel) gives
			\begin{align}
			 0 &= \Bra{b_i}[\p{K}{-1}_1, \p{H}{0}]\Ket{b_j}\\
			 	\notag &= (1-\delta_{j,0})\,[[\p{K}{-1}_1]]_{i,\,j-1}
			 - (1-\delta_{i,0})\,[[\p{K}{-1}_1]]_{i-1,\,j}\,.
			 \label{eq:comm_proj}
			\end{align}
			For $i,j \neq 0$ this reads
			\begin{align}
			 [[\p{K}{-1}_1]]_{i,\,j-1} = [[\p{K}{-1}_1]]_{i-1,\,j}\,.
			\end{align}
			In the Jordan basis, $[[\p{K}{-1}_1]]_{ij}$ corresponds to the $(N-1-i,\,j)$-th element of the matrix representation of $\p{K}{-1}_1$\,, which we denote by $M$\,, i.e.\ $[[\p{K}{-1}_1]]_{ij} = M_{N-1-i,\,j}$\,. The relation above then implies
			\begin{align}
			 M_{N-1-i,\,j-1} = M_{N-i,\,j}\,, \qquad i,j \neq 0\,,
			\end{align}
			which means every element along each diagonal of $M$ is the same; $M$ is therefore a Toeplitz matrix. Moreover, by Eq.~\eqref{eq:K_leading},
			\begin{align}
			 [[\p{K}{-1}_1]]_{k,0} = M_{N-1-k,\,0} = 0\,,
			 \qquad k = 0,1,\ldots,N-1\,,
			\end{align}
			so that the entire first column of $M$ vanishes. Together with the Toeplitz property, this sets every entry of $M_{ij}$ with $i - j \geq 0$ to zero. Thus $M$ is a strictly upper-triangular Toeplitz matrix. In terms of $[[\p{K}{-1}_1]]$\,, we have
			\begin{align}
				\label{Km1}[[\p{K}{-1}_1]]_{N-1-i, j}=0\,,~~~i-j\geq 0\,.
			\end{align}
			In addition to the zero entries, we can determine the non-zero entries of $\p{K}{-1}_1$ necessary in the main text. Taking Eq.~\eqref{eq:K1proj} with $s = -N+1$\,, the left-hand side receives a contribution only from $(m,r') = (-1,1)$\,, and the right-hand side only from $(r,r') = (1,0)$\,, giving
			\begin{align}
			 [[\p{K}{-1}_1]]_{N-1,1} = \frac1N\,.
			\end{align}
			Through the Toeplitz property, this means
			\begin{align}
			 [[\p{K}{-1}_1]]_{1,N-1} = \frac1N\,.
			 \label{eq:K1m1_super}
			\end{align}
			This is crucial for the $\p{h}{2/N}$ computation.

	\section{Equivalence of the Genericity Condition}
		\label{app:generic}
		The generic condition introduced by Welters~\cite{Welters2011} involves the $\epsilon$-derivative of the characteristic polynomial $\det\!\left(z I - H(\epsilon)\right)$ to be non-vanishing at $(\epsilon,z)=(0,\p{h}{0})$\,. We show that, for arbitrary $N$, it evaluates to
\begin{align}
    \left.\frac{\partial}{\partial\epsilon}\det\!\left(z I - H(\epsilon)\right)\right|_{(\epsilon,z)=(0,\p{h}{0})} = -[[\p{H}{1}]]_{00} \neq 0\,.
    \label{eq:welters_equiv}
\end{align}
Under this condition, the $N$ eigenvalues near a pure order-$N$ EP are the $N$ branches of a single convergent Puiseux series in $\epsilon^{1/N}$\,. Moreover, the nonvanishing required by the generic condition is precisely the condition $\left(\p{h}{1/N}\right)^N = [[\p{H}{1}]]_{00} \neq 0$ of Eq.~\eqref{h1N}.

To clearly establish Eq.~\eqref{eq:welters_equiv}, we define $P(\epsilon,z) = z I - \p{H}{0} - \epsilon\,\p{H}{1}$\,. Using Jacobi's formula, the derivative of the determinant is given by $\partial_\epsilon \det P = \operatorname{tr}\!\left(\operatorname{adj}(P)\,\partial_\epsilon P\right)$\,. At $(\epsilon,z)=(0,\p{h}{0})$\,, we have $\partial_\epsilon P = -\p{H}{1}$ and $P(0,\p{h}{0}) = \p{h}{0} I - \p{H}{0}$\,. Because the unperturbed Hamiltonian $\p{H}{0}$ is an upper Jordan block, with $\p{h}{0}$ on the diagonal and unit entries on the superdiagonal, this zero-order matrix reduces to $-Q$\,, where $Q$ is the nilpotent shift matrix defined by $Q_{i,i+1}=1$ and zero elsewhere. Evaluating the derivative therefore yields
\begin{align}
    \left.\frac{\partial}{\partial\epsilon}\det P\right|_{(0,\p{h}{0})} = -\operatorname{tr}\!\left(\operatorname{adj}(-Q)\,\p{H}{1}\right)\,.
\end{align}

		Since the shift matrix $Q$ has rank $N-1$, its adjugate $\operatorname{adj}(-Q)$ has rank one. The $(i,j)$ entry of this adjugate is the signed $(j,i)$ cofactor of $-Q$, which is non-zero if and only if we delete the identically-zero last row ($j = N-1$) and first column ($i = 0$) of $-Q$. Deleting these leaves an $(N-1)\times(N-1)$ upper-triangular submatrix with $-1$ on its diagonal, yielding a determinant of $(-1)^{N-1}$. Multiplying this by the cofactor sign $(-1)^{N-1}$ gives exactly $(-1)^{2(N-1)} = 1$. Therefore, $\operatorname{adj}(-Q)$ is simply a matrix with a single $1$ in the $(0,N-1)$ entry and zeros elsewhere. Taking the trace of this matrix multiplied by $\p{H}{1}$ precisely extracts the $(N-1,0)$ element of $\p{H}{1}$, reducing the derivative to
\begin{align}
\left.\frac{\partial}{\partial\epsilon}\det P\right|_{(0,\p{h}{0})}
= -\left(\p{H}{1}\right)_{N-1,0}
= -[[\p{H}{1}]]_{00}\,.
\end{align}
This completes the proof of Eq.~\eqref{eq:welters_equiv}. We emphasize that this result depends only on the single matrix element $[[\p{H}{1}]]_{00}$, independently of whether the perturbation $\p{H}{1}$ is invertible. Although the computation is carried out in the explicit Jordan form~\eqref{JordanForm}, the result is representation independent: both $\det(zI - H(\epsilon))$ and $[[\p{H}{1}]]_{00}$ are invariant under a change of basis, so that Eq.~\eqref{eq:welters_equiv} holds in any basis. The representation~\eqref{JordanForm} serves only for computational convenience; the proof may equally be carried out using the Jordan-chain and biorthogonality relations~\eqref{JCR}--\eqref{COMP} alone.
	\end{appendix}

\bibliography{References}
\end{document}